\documentclass[11pt]{article}

\usepackage{latexsym,amsmath}

\newcounter{subequation}[equation]
\makeatletter

\def\bcite{\@ifnextchar [{\@tempswatrue\@bcitex}{\@tempswafalse\@bcitex[]}}
\def\@bcitex[#1]#2{\if@filesw\immediate\write\@auxout{\string\citation{#2}}\fi
  \let\@bcitea\@empty
  \@bcite{\@for\@bciteb:=#2\do
    {\@bcitea\def\@bcitea{,\penalty\@m\ }%
     \def\@tempa##1##2\@nil{\edef\@bciteb{\if##1\space##2\else##1##2\fi}}%
     \expandafter\@tempa\@bciteb\@nil
     \@ifundefined{b@\@bciteb}{{\reset@font\bf ?}\@warning
       {Citation `\@bciteb' on page \thepage \space undefined}}%
     \hbox{\csname b@\@bciteb\endcsname}}}{#1}}
\def\@bcite#1#2{{#1\if@tempswa , #2\fi}}

\def\thesubequation{\theequation\@alph\c@subequation}
\def\@subeqnnum{{\rm (\thesubequation)}}
\def\slabel#1{\@bsphack\if@filesw {\let\thepage\relax
   \xdef\@gtempa{\write\@auxout{\string
      \newlabel{#1}{{\thesubequation}{\thepage}}}}}\@gtempa
   \if@nobreak \ifvmode\nobreak\fi\fi\fi\@esphack}
\def\subeqnarray{\stepcounter{equation}
\let\@currentlabel=\theequation\global\c@subequation\@ne
\global\@eqnswtrue
\global\@eqcnt\z@\tabskip\@centering\let\\=\@subeqncr
$$\halign to \displaywidth\bgroup\@eqnsel\hskip\@centering
  $\displaystyle\tabskip\z@{##}$&\global\@eqcnt\@ne
  \hskip 2\arraycolsep \hfil${##}$\hfil
  &\global\@eqcnt\tw@ \hskip 2\arraycolsep
  $\displaystyle\tabskip\z@{##}$\hfil
   \tabskip\@centering&\llap{##}\tabskip\z@\cr}
\def\endsubeqnarray{\@@subeqncr\egroup
                     $$\global\@ignoretrue}
\def\@subeqncr{{\ifnum0=`}\fi\@ifstar{\global\@eqpen\@M
    \@ysubeqncr}{\global\@eqpen\interdisplaylinepenalty \@ysubeqncr}}
\def\@ysubeqncr{\@ifnextchar [{\@xsubeqncr}{\@xsubeqncr[\z@]}}
\def\@xsubeqncr[#1]{\ifnum0=`{\fi}\@@subeqncr
   \noalign{\penalty\@eqpen\vskip\jot\vskip #1\relax}}
\def\@@subeqncr{\let\@tempa\relax
    \ifcase\@eqcnt \def\@tempa{& & &}\or \def\@tempa{& &}
      \else \def\@tempa{&}\fi
     \@tempa \if@eqnsw\@subeqnnum\refstepcounter{subequation}\fi
     \global\@eqnswtrue\global\@eqcnt\z@\cr}
\let\@ssubeqncr=\@subeqncr
\@namedef{subeqnarray*}{\def\@subeqncr{\nonumber\@ssubeqncr}\subeqnarray}
\@namedef{endsubeqnarray*}{\global\advance\c@equation\m@ne%
                           \nonumber\endsubeqnarray}

\makeatother

\DeclareFontFamily{OT1}{rsfs10}{}
\DeclareFontShape{OT1}{rsfs10}{m}{n}{ <-> rsfs10 }{}
\DeclareMathAlphabet{\mathscript}{OT1}{rsfs10}{m}{n}

\numberwithin{equation}{section}

\newcommand{\ns}{\normalsize}

\newcommand{\be}{\begin{equation}}
\newcommand{\ee}{\end{equation}}
\newcommand{\nn}{\nonumber}
\newcommand{\bea}{\begin{eqnarray}}
\newcommand{\eea}{\end{eqnarray}}
\newcommand{\bsea}{\begin{subeqnarray}} 
\newcommand{\esea}{\end{subeqnarray}}

\newcommand{\tr}{\textrm{tr}}

\def\a{\alpha}
\def\b{\beta}
\def\g{\gamma}

\def\d{\delta}
\def\e{\epsilon}

\def\z{\psi}

\def\k{\kappa}
\def\l{\lambda}
\def\m{\mu}
\def\n{\nu}
\def\o{\omega}
\def\p{\pi}

\def\r{\rho}
\def\s{\sigma}

\def\t{\tau}

\def\x{\xi}
\def\z{\zeta}
\def\w{\wedge}

\def\G{\Gamma}
\def\J{\Psi}

\def\O{\Omega}

\def\cA{{\cal A}}
\def\cF{{\cal F}}
\def\cM{{\cal M}}

\begin{document}


\begin{titlepage}

\vspace{-2cm}

\title{
   \hfill{\ns UPR-797T, PUPT-1782\\}
   \hfill{\ns CERN-TH/98-98, Imperial/TP/97-98/35\\}
   \hfill{\ns hep-th/9803235\\[.5cm]}
   {\LARGE The Universe as a Domain Wall}}
\author{
   Andr\'e Lukas$^1$
      \setcounter{footnote}{0}\thanks{Supported in part by Deutsche
          Forschungsgemeinschaft (DFG).}~~,
   Burt A.~Ovrut$^1$
      \setcounter{footnote}{3}\thanks{Supported in part by a Senior 
          Alexander von Humboldt Award.}~~,
   K.S. Stelle$^2$ and Daniel Waldram$^3$\\[0.5cm]
   {\ns $^1$Department of Physics, University of Pennsylvania} \\
   {\ns Philadelphia, PA 19104--6396, USA}\\[0.3cm]
   {\ns $^2$The Blackett Laboratory, Imperial College, London SW7 2BZ, UK}\\ 
   {\ns and}\\
   {\ns TH Division, CERN, CH-1211 Geneva 23, Switzerland}\\[0.3cm]
   {\ns $^3$Department of Physics, Joseph Henry Laboratories,}\\ 
   {\ns Princeton University, Princeton, NJ 08544, USA}}
\date{}

\maketitle

\begin{abstract}

It is shown that the effective five--dimensional
theory of the strongly coupled heterotic string is a gauged version
of $N=1$ five--dimensional supergravity with four--dimensional boundaries. 
For the universal supermultiplets, this theory is explicitly 
constructed by a generalized dimensional reduction procedure on a Calabi-Yau
manifold. A crucial ingredient in the reduction is the retention of a
``non--zero mode'' of the four--form field strength, leading to the gauging
of the universal hypermultiplet by the graviphoton. We show that this theory
has an exact three--brane domain wall solution which reduces to Witten's
``deformed'' Calabi--Yau background upon linearization. This solution
consists of two parallel three--branes with sources provided by the
four--dimensional boundary theories and constitutes the appropriate background
for a reduction to four dimensions. Four--dimensional space--time is then
identified with the three--brane worldvolume. 
\end{abstract}

\thispagestyle{empty}

\end{titlepage}


\section{Introduction}


The strongly coupled $E_8\times E_8$ heterotic string has been
identified as the eleven-dimensional limit of M--theory
compactified on an $S^1/Z_2$ orbifold with a set of $E_8$ gauge fields
at each ten-dimensional orbifold fixed plane~\cite{hw1,hw2}. Witten
has shown that there exists a consistent compactification of this M--theory
limit on a deformed Calabi-Yau threefold, leading to a supersymmetric $N=1$
theory in four dimensions~\cite{w}. Matching at tree level to the
phenomenological gravitational and grand-unified couplings~\cite{w,bd}, one
finds the orbifold must be larger than the Calabi-Yau radius, which is of
the order of the eleven-dimensional Planck length. This suggests that there
is a regime where the universe appears five-dimensional. It is
then important to find the five--dimensional effective action,
describing the low-energy physics of the strongly coupled heterotic
string and which underlies phenomenologically relevant
four--dimensional $N=1$ supergravity models. 
Furthermore, this theory constitutes a new setting for early universe string
(M--theory) cosmology, which has traditionally been studied in the framework
of the four--dimensional effective action. Although some formal and
phenomenological aspects of the strongly coupled heterotic string have been
studied in the literature~[\bcite{hor}\,--\,\bcite{bkl}], a derivation of the
five--dimensional effective action from Ho\v rava--Witten theory, and a
detailed discussion of its properties, have remained missing. (Some aspects of
five--dimensional physics, however, were considered in
\cite{bd,noy,sharpe,pes}.)

In the present paper, we derive this effective five-dimensional theory for
the universal bulk fields; that is, the gravity supermultiplet and the
universal hypermultiplet. We shall show that the relevant consistent reduction
from eleven to five dimensions on a Calabi-Yau manifold requires the inclusion
of non-zero values of the 4--form field strength in the internal Calabi-Yau
directions. This leads to a gauged five-dimensional supergravity action
with a potential term that has not previously been constructed. More precisely,
given the universal hypermultiplet coset manifold~\cite{quat}
$\cM_Q=SU(2,1)/SU(2)\times U(1)$, we find that a subgroup
$U(1)\subset SU(2)\times U(1)$ is gauged, with the vector field in
the gravity supermultiplet as the corresponding gauge boson. Owing to the
potential, flat space is not a solution of this five--dimensional
theory without the Calabi-Yau space decompactifying. However, the
equations of motion do admit a three-brane solution that preserves
half of the remaining $D=5$ supersymmetries where the Calabi-Yau
remains compact.
This is supported by source terms on the fixed orbifold planes
of the five-dimensional space. This BPS three--brane constitutes the
``vacuum'' of the five--dimensional theory and it is the appropriate
background for a further reduction to four--dimensional $N=1$ supergravity
theories. In such a reduction, four--dimensional space--time becomes
identified with the three--brane worldvolume. We will show that
the linearized version of this three--brane corresponds to Witten's
``deformed'' Calabi-Yau solution, which was constructed only to first
non--trivial order in powers of the eleven--dimensional Newton constant.
Thus, our solution represents a generalization of this original background,
as it is an exact solution of the effective low energy theory. The inversion
of the Ho\v{r}ava-Witten construction by first performing a generalized
Kaluza-Klein reduction from eleven down to five dimensions, and then finally
from five to four dimensions, is more natural for two reasons. First,
as noted above, the
scale of the fifth dimension is larger than that of the Calabi-Yau
manifold. Secondly, the generalized Kaluza-Klein reduction is a consistent
truncation, meaning that, from the point of view of the bulk theory,
the heavy Calabi-Yau modes can simply be consistently 
set to zero without inducing higher-order corrections. The reduction from
five to four dimensions will, however, require carefully integrating out
the non-trivial five-dimensional modes, giving rise to higher-order
corrections of potential
phenomenological interest. The relation between Witten's deformed
Calabi-Yau solution and the five--dimensional domain wall solution can also be
described using brane language. As we will see, there is a natural
interpretation of Witten's solution as a collection of five-branes wrapped on
two-cycles of the Calabi-Yau space and lying in the orbifold fixed planes.
Reduced to five dimensions, these then become three-branes spanning the
orbifold fixed planes. 

In summary, we argue that it is a gauged version of five-dimensional
supergravity that is the correct arena for considering the effective action of
the strongly coupled $E_8\times E_8$ heterotic string in the intermediate
energy range. This effective theory has three--brane domain--wall BPS
solutions, with the three--brane worldvolume corresponding to the orbifold
planes. These solutions represent the correct background for making contact
with four--dimensional low--energy physics.

Let us now summarize our conventions. We will
consider eleven-dimensional spacetime compactified on a Calabi-Yau space $X$,
with the subsequent reduction down to four dimensions effectively provided by
a double-domain-wall background, corresponding to an $S^1/Z_2$ orbifold. We
use coordinates $x^{I}$ with indices $I,J,K,\cdots = 0,\cdots ,9,11$ to
parameterize the full 11--dimensional space $M_{11}$. Throughout this paper,
when we refer to orbifolds, we will work in the ``upstairs'' picture
with the orbifold $S^1/Z_2$ in the $x^{11}$--direction. We choose the range
$x^{11}\in [-\pi\rho ,\pi\rho ]$ with the endpoints being identified. The
$Z_2$ orbifold symmetry acts as $x^{11}\rightarrow -x^{11}$. Then there exist
two ten--dimensional hyperplanes fixed under the $Z_2$ symmetry which we
denote by $M_{10}^{(i)}$, $i=1,2$. Locally, they are specified by the
conditions $x^{11}=0,\pi\rho$. Barred indices
$\bar{I},\bar{J},\bar{K},\cdots = 0,\cdots ,9$ are used for the
ten--dimensional space orthogonal to the orbifold. 
Upon reduction on the Calabi-Yau space we have a five-dimensional spacetime
$M_5$ labeled by indices $\a ,\b ,\g ,\cdots  = 0,\cdots ,3,11$. The
orbifold fixed planes become four-dimensional with indices
$\m,\n,\rho,\cdots = 0,\cdots ,3$. We use indices $A,B,C,\cdots =
4,\cdots 9$ for the Calabi--Yau space. 
The 11-dimensional Dirac--matrices $\G^I$ with $\{\G^I,\G^J\}=2g^{IJ}$
are decomposed as $\G^I = \{\g^\a\otimes\l ,{\bf 1}\otimes\l^A\}$
where $\g^\a$ and $\l^A$ are the five-- and six--dimensional Dirac
matrices, respectively. Here, $\l$ is the chiral projection matrix in
six dimensions with $\l^2=1$. Spinors in eleven dimensions will be
Majorana spinors with 32 real components throughout the paper. In five
dimensions we use symplectic-real spinors~\cite{c0} $\psi^i$ where
$i=1,2$ is an $SU(2)$ index, corresponding to the automorphism
group of the $N=1$ supersymmetry algebra in five dimensions. We will
follow the conventions given in~\cite{GST1}.
Fields will be required to have a definite behaviour under the $Z_2$
orbifold symmetry in $D=11$. We demand a bosonic field $\Phi$ to be
even or odd; that is, $\Phi (x^{11})=\pm\Phi (-x^{11})$. For a spinor
$\Psi$ the condition is $\G_{11}\Psi (-x^{11})=\Psi (x^{11})$ so that
the projection to one of the orbifold planes leads to a
ten--dimensional Majorana--Weyl spinor with positive
chirality. Similarly, in five dimensions, bosonic fields will be
either even or odd. We can choose a basis for the $SU(2)$ automorphism
group such that symplectic-real spinors $\psi^i$ 
satisfy the constraint
$\g_{11}\psi^i(-x^{11})=(\t_3)^i_j\psi^j(x^{11})$ where $\t_a$ are the
Pauli spin matrices, so $\t_3=\mbox{diag}(1,-1)$.


\section{The strongly coupled heterotic string and Calabi--Yau solutions}

To set the scene for our later discussion, we will now briefly review the
effective description of strongly coupled heterotic string theory as
11-dimensional supergravity with boundaries given by Ho\v{r}ava and
Witten~\cite{hw1,hw2}. In addition, we present, in a simple form, the
solutions of this theory~\cite{w} appropriate for a reduction to $N=1$
theories in four dimensions using the explicit form of these solutions given
in ref.~\cite{low1}.

The bosonic part of the action is of the form
\begin{equation}
\label{action}
   S = S_{\rm SG}+S_{\rm YM}
\end{equation}
where $S_{\rm SG}$ is the familiar 11--dimensional supergravity
\begin{equation}
 S_{\rm SG} = -\frac{1}{2\k^2}\int_{M^{11}}\sqrt{-g}\left[ 
                    R+\frac{1}{24}G_{IJKL}G^{IJKL}
           +\frac{\sqrt{2}}{1728}\e^{I_1...I_{11}}
               C_{I_1I_2I_3}G_{I_4...I_7}G_{I_8...I_{11}} \right]
 \label{SSG}
\end{equation}
and $S_{\rm YM}$ are the two $E_8$ Yang--Mills theories on the orbifold planes
explicitly given by~\footnote{We note that there is a debate in the
literature about the precise value of the Yang--Mills coupling constant
in terms of $\k$. While we quote the original value~\cite{hw2,deA} the
value found in ref.~\cite{conrad} is smaller.
In the second case, the coefficients in the Yang-Mills action~\eqref{SYM} 
and the Bianchi identity~\eqref{Bianchi} should both be multiplied by 
$2^{-1/3}$. This potential factor will not be essential in the following 
discussion as it will simply lead to a redefinition of the five--dimensional 
coupling constants. We will comment on this point later on.}
\begin{multline}
   \label{SYM}
   S_{\rm YM} = - \frac{1}{8\pi\k^2}\left(\frac{\k}{4\pi}\right)^{2/3}
        \int_{M_{10}^{(1)}}\sqrt{-g}\;\left\{
           \tr(F^{(1)})^2 - \frac{1}{2}\tr R^2\right\} \\
        - \frac{1}{8\pi\k^2}\left(\frac{\k}{4\pi}\right)^{2/3}
           \int_{M_{10}^{(2)}}\sqrt{-g}\;\left\{
               \tr(F^{(2)})^2 - \frac{1}{2}\tr R^2\right\}\; .
\end{multline}
Here $F_{\bar{I}\bar{J}}^{(i)}$ are the two $E_8$ gauge field strengths and
$C_{IJK}$ is the 3--form with field strength
$G_{IJKL}=24\,\partial_{[I}C_{JKL]}$. In order for the above 
theory to be supersymmetric as well as anomaly free, the Bianchi
identity for $G$ should receive a correction such that
\begin{equation}
 (dG)_{11\bar{I}\bar{J}\bar{K}\bar{L}} = -\frac{1}{2\sqrt{2}\pi}
    \left(\frac{\k}{4\pi}\right)^{2/3} \left\{ 
       J^{(1)}\d (x^{11}) + J^{(2)}\d (x^{11}-\pi\r )
       \right\}_{\bar{I}\bar{J}\bar{K}\bar{L}} \label{Bianchi}
\end{equation}
where the sources are given by 
\begin{equation}
 J^{(i)}
    = \left( {\rm tr}F^{(i)}\wedge F^{(i)} 
      - \frac{1}{2}{\rm tr}R\wedge R \right)\; .\label{J}
\end{equation}
Under the $Z_2$ orbifold symmetry, the field components $g_{\bar{I}\bar{J}}$,
$g_{11,11}$, $C_{\bar{I}\bar{J}11}$ are even, while $g_{\bar{I}11}$,
$C_{\bar{I}\bar{J}\bar{K}}$ are odd. We note that the above boundary
actions contain, in addition to the Yang--Mills terms, $\tr R^2$ terms which
were not part of the original theory
derived in~\cite{hw2}. It was argued in ref.~\cite{low1} that these terms
are required by supersymmetry, since they pair with the $R^2$ terms in the
Bianchi identity~\eqref{Bianchi} in analogy to the weakly coupled case.
The existence of these terms will be of some importance in the following.

One way to view this theory is to draw an analogy between the orbifold
planes and D--branes in type II theories. A collection of D$p$-branes
is described by a $U(N)$ gauge theory. The D$p$-brane charge is
measured by $\tr{\bf 1}=N$, while exciting a D$(p-2)$-brane charge
corresponds to having a non-trivial $\tr F$, and a D$(p-4)$-brane
charge corresponds to non-trivial $\tr F\w F$ and
so on~\cite{li1}. Similarly, if the original D-branes are on a curved
manifold then there is also an induced charge for lower-dimensional
branes given by $\tr R\w R$ and higher even powers~\cite{ghm}.
Applying this picture to our situation,  the r\^ole of the $U(N)$ gauge
field on the $D$--brane worldvolume is here played by the $E_8$ gauge fields
on the orbifold planes. The correction to the Bianchi
identity then has the interpretation of exciting an M5-brane charge in
the orbifold plane. In ref.~\cite{llo} this picture has been made explicit by
constructing a gauge five--brane in this theory.

We would now like to discuss solutions of the above theory which preserve
four of the 32 supercharges leading, upon compactification, to four
dimensional $N=1$ supergravities. This task is significantly complicated by
the fact that the sources in the Bianchi identity~\eqref{Bianchi} are
located on the orbifold planes with the gravitational part distributed
equally between the two planes. While the standard embedding of the
spin connection into the gauge connection
\begin{equation}
 \tr F^{(1)}\w F^{(1)} = \tr R\w R
\label{condition}
\end{equation}
leads to vanishing source terms in the weakly coupled heterotic string
Bianchi identity (which, in turn, allows one to set the antisymmetric tensor
gauge field to zero), in the present case, one is left with non--zero sources
$\pm\tr R\w R$ on the two hyperplanes. As a result, the antisymmetric tensor
field
$G$ and, hence, the second term in the gravitino supersymmetry variation
\begin{equation} 
 \d\J_I = D_I\eta +\frac{\sqrt{2}}{288}\left(\G_{IJKLM}-8g_{IJ}\G_{KLM}
          \right)G^{JKLM}\eta +\; \cdots \label{susy}\; ,
\end{equation}
do not vanish. Thus, straightforwardly compactifying on a Calabi--Yau manifold
no longer provides a solution to the Killing spinor equation $\d\J_I =0$.
The problem can, however, be treated perturbatively in powers
of the 11--dimensional 
Newton constant $\k$. To lowest order, one can start with a manifold
$X\times S^1/Z_2\times M_4$ where $X$ is a Calabi--Yau three--fold and
$M_4$ is four--dimensional Minkowski space. This manifold has an
$x^{11}$--independent (and hence chiral) Killing spinor $\eta$ which
corresponds to four preserved supercharges. Then, one can determine the 
first order corrections to this background and the spinor $\eta$ so that
the gravitino variation vanishes to order $\k^{2/3}$.

The existence of such a distorted background solution to order
$\k^{2/3}$  has been demonstrated in ref.~\cite{w}. To see its explicit
form, let us start with the zeroth order metric
\begin{equation}
 ds_{11}^2 = \eta_{\m\n}dx^\m dx^\n+R_0^2(dx^{11})^2+V_0^{1/3}\O_{AB}dx^Adx^B
             \; ,
\end{equation}
where $\O_{AB}$ is a Calabi--Yau metric with K\"ahler form
$\o_{a\bar{b}}=i\O_{a\bar{b}}$. (Here $a$ and $\bar{b}$ are holomorphic
and anti-holomorphic indices.) To keep track of the scaling properties of 
the solution, we have introduced moduli $V_0$ and $R_0$ for the Calabi--Yau
volume and the orbifold radius, respectively. It was shown in~\cite{w}
that, to order $\k^{2/3}$, the metric can be written in the form
\begin{equation}
 ds_{11}^2 = (1+\hat{b})\eta_{\m\n}dx^\m dx^\n+R_0^2(1+\hat{\g})(dx^{11})^2
             +V_0^{1/3}(\O_{AB}+h_{AB})dx^Adx^B \label{metric}
\end{equation}
where the functions $\hat{b}$, $\hat{\g}$ and $h_{AB}$ depend on $x^{11}$
and the Calabi--Yau coordinates. Furthermore, as we have discussed, $G_{ABCD}$
and $G_{ABC11}$ receive a contribution of order $\k^{2/3}$ from the Bianchi
identity source terms. To get the general explicit form of the corrections,
one has to solve the relations given in ref.~\cite{w}. This can be done by
dualizing the antisymmetric tensor field and using a harmonic expansion on the
Calabi--Yau space~\cite{low1}. 

Here, we quote those results simplified in two
essential ways. First, we drop all terms corresponding to non--zero
eigenvalue harmonics on the Calabi--Yau space. These terms will be of no
relevance to the low energy theory, since they correspond to heavy
Calabi--Yau modes which decouple at this order. Second, we write only
the one massless term that is related to the Calabi--Yau breathing
mode. This will be sufficient for all applications dealing only with the
universal moduli. Given these simplifications, the corrections are
explicitly 
\bsea
 \hat{b} &=& -\frac{\sqrt{2}}{3}R_0V_0^{-2/3}\a\, (|x^{11}|-\p\r /2)\\
 \hat{\g} &=& \frac{2\sqrt{2}}{3}R_0V_0^{-2/3}\a\, (|x^{11}|-\p\r /2)\\
 h_{AB} &=& \frac{\sqrt{2}}{3}R_0V_0^{-2/3}\a\, (|x^{11}|-\p\r /2)\O_{AB}\\
 G_{ABCD} &=& \frac{1}{6}\a\,{\e_{ABCD}}^{EF}\,\o_{EF}\,\e (x^{11})\\
 G_{ABC11} &=& 0\label{corr} 
\esea
with
\begin{equation}
 \a = -\frac{1}{8\sqrt{2}\p v}\left(\frac{\k}{4\p}\right)^{2/3}\int_X
      \o\wedge\tr R^{(\O )}\wedge R^{(\O )}\; ,\qquad
 v=\int_X\sqrt{\O}\; .\label{alpha}
\end{equation}
Here $\e (x^{11})$ is the step function which is $+1$ ($-1$) for $x^{11}$
positive (negative).
Note that, by dropping the massive modes, these expressions take
a very simple form representing a linear increase of the corrections along
the orbifold. Even more significantly, and unlike the exact solution
including the heavy modes, the above approximation leads to a corrected metric
$\O_{AB}+h_{AB}$ that is still of Calabi--Yau type  at each point on the 
$S^1/Z_2$ orbifold. The Calabi--Yau volume (and, if all moduli are included,
also its shape), however, is continuously changing across the orbifold. More
generally, one can think of the internal part of the corrected metric as a
curve in the Calabi--Yau moduli space.

Returning to the D--brane perspective, one can view the above configuration as
the linearized solution for a collection of five-branes embedded in the
orbifold planes. The relation~\eqref{condition} fixes equal amounts of
five-brane charge, $\frac{1}{2}\tr R\w R$, on each orbifold fixed plane, where
the five-branes are confined to live. Since $\tr R\w R \in
H^{2,2}(X)$, we can associate a different five-brane charge for each
independent element of $H^{2,2}(X)$. The five--branes themselves are
associated with Poincar\'e dual cycles. Thus they span the non--compact
four-dimensional space together with a two-cycle in the Calabi-Yau space.
In particular, from the five-dimensional point of view, they are three-branes
localized on the orbifold planes. Witten's construction ensures that this
configuration of branes preserves one-eighth of the supersymmetry. Finally,
restricting to just the Calabi-Yau breathing modes corresponds to keeping only
the five--brane which spans the holomorphic two-cycle in the
Calabi-Yau defined by the K\"ahler form. 


\section{The five--dimensional effective action}


Phenomenologically, there is a regime where the universe appears
five-dimensional. We would, therefore, like to derive an effective
theory in the space consisting of the usual four space-time dimensions
and the orbifold, based on the background solution discussed in the
previous section. As we have already mentioned, we will consider the
universal zero modes only; that is, the five--dimensional graviton
supermultiplet and the breathing mode of the Calabi--Yau space, along
with its superpartners. These form a hypermultiplet in five
dimensions. Furthermore, to keep the discussion as simple as possible,
we will not consider boundary gauge matter fields. This simple
framework suffices to illustrate our main ideas. The general case will
be presented elsewhere~\cite{losw}. 

Naively, one might attempt to perform the actual reduction directly on
the background given in eqs.~\eqref{metric} and \eqref{corr}.
This would, however, lead to a complicated five--dimensional theory with
explicit $x^{11}$--dependence in the action. Moreover, this background
preserves only four supercharges whereas the minimal supergravity in
five dimensions ($N=1$) is invariant under twice this amount of supersymmetry.

A useful observation here is that, since we retain the dependence on the
orbifold coordinate, we can actually absorb the metric deformations in
\eqref{metric} and \eqref{corr} into the five--dimensional metric moduli. That
is, the $x^{11}$--dependent scale factors $\hat{b}$ and $\hat{\g}$ of the
four--dimensional space and of the orbifold can be absorbed into the
five--dimensional (Einstein frame) metric $g_{\a\b}$ while,
analogously, the variation of the Calabi--Yau volume along the orbifold
encoded in $h_{AB}$ can be absorbed into a modulus
$V$.~\footnote{Note that we could not apply a similar method for a
reduction down to four dimensions, as all moduli fields would then be
$x^{11}$ independent. In this case, one should work with the
background in the form \eqref{metric}, \eqref{corr} as done in
ref.~\cite{low1}.} More precisely, we can perform the Kaluza-Klein
reduction on the metric 
\begin{equation}
 ds_{11}^2 = V^{-2/3}g_{\a\b}dx^\a dx^\b +V^{1/3}\O_{AB}dx^Adx^B\; .
 \label{metric1}
\end{equation}
This rewriting suggests a change of perspective: rather than reducing on the
Witten vacuum, we can try to find an effective five-dimensional theory
where we recover the Witten vacuum as a particular solution.

We see that, since we have absorbed the deformation into the
moduli, the background corresponding to the metric~\eqref{metric1}
preserves eight supercharges, the appropriate number for a reduction down to
five dimensions. It might appear that we are
simply performing a standard reduction of 11--dimensional supergravity
on a Calabi--Yau space to five dimensions; for example, in the way described in
ref.~\cite{CYred}. If this were the case, then it would be hard to understand
how the resulting five--dimensional theory could encode any information about
the deformed Calabi--Yau background. There are, however, two important
ingredients that we have not yet included. One is obviously the existence of
the boundary theories. We will return to this point shortly. First, however,
let us explain a somewhat unconventional addition to the bulk theory that
must be included.

Although we could absorb all metric corrections into the five--dimensional
metric moduli, the same is not true for the 4--form
field. Specifically, for the nonvanishing component $G_{ABCD}$ in
eq.~(\ref{corr}d) there is no corresponding zero mode
field~\footnote{This can be seen from the mixed part of the Bianchi
identity $\partial_\a G_{ABCD}=0$ which shows that the constant $\a$
in eq.~\eqref{corr} cannot be promoted as stands to a five--dimensional
field. It is possible to dualize in five dimensions so the constant
$\a$ is promoted to a five-form field, but we will not pursue this
formulation here.}.
Therefore, in the reduction, we should take this part of $G$ explicitly
into account. In the terminology of ref.~\cite{gsw}, such an antisymmetric
tensor field configuration is called a ``non--zero mode''. More generally, a
non--zero mode is a background antisymmetric tensor field that solves the
equations of motion but, unlike antisymmetric tensor field moduli, has
nonvanishing field strength. Such configurations, for a $p$--form field
strength, can be identified with the cohomology group $H^p(M)$ of the manifold
$M$ and, in particular, exist if this cohomology group is nontrivial. In the
case under consideration, the relevant cohomology group is $H^4(X)$ which
is nontrivial for a Calabi--Yau manifold $X$ since $h^{2,2}=h^{1,1}\geq 1$.
Again, the form of $G_{ABCD}$ in eq.~(\ref{corr}d) is somewhat special,
reflecting the fact that we are concentrating here on the universal moduli. In
the general case, $G_{ABCD}$ would be a linear combination of all
harmonic $(2,2)$--forms.

The complete configuration for the antisymmetric tensor field that we use in
the reduction is given by
\bea
 C_{\a\b\g}&,\quad&G_{\a\b\g\d}=24\,\partial_{[\a}C_{\b\g\d]}\nn \\
 C_{\a AB} = \frac{1}{6}\cA_\a\o_{AB}&,\quad&G_{\a\b AB}=\cF_{\a\b}\o_{AB}\; ,
             \quad \cF_{\a\b}=\partial_\a\cA_\b -\partial_\b\cA_\a
              \label{Gmod}\\
 C_{ABC} = \frac{1}{6}\x\o_{ABC}&,\quad&G_{\a ABC}=\partial_\a\x\o_{ABC} \nn
\eea
and the non--zero mode is
\begin{equation}
 G_{ABCD} = \frac{\a}{6}{\e_{ABCD}}^{EF}\,\o_{EF}\,\epsilon (x^{11})\; ,
            \label{nonzero}
\end{equation}
where $\a$ was defined in eq.~\eqref{alpha}. Here, $\o_{ABC}$ is the
harmonic $(3,0)$ form on the Calabi--Yau space and $\x$ is the corresponding
(complex) scalar zero mode. In addition, we have a five-dimensional vector
field $\cA_\a$ and 3--form $C_{\a\b\g}$, which can be
dualized to a scalar $\s$. The total bulk field content of the
five--dimensional theory is then given by the gravity multiplet
$(g_{\a\b},\cA_\a ,\psi^i_\a )$ together with the universal hypermultiplet
$(V,\s ,\x ,\bar{\x},\z^i)$ where $\psi_\a^i$ and
$\z^i$ are the gravitini and the hypermultiplet fermions respectively and
$i=1,2$. From their relations to the 11--dimensional fields, it is easy to see
that $g_{\m\n}$, $g_{11,11}$, $\cA_{11}$, $\s$ must be even under the
$Z_2$ action whereas $g_{\m 11}$, $\cA_\m$, $\x$ must be odd.

Examples of compactifications with non--zero modes in pure 11--dimensional
supergravity on various manifolds including Calabi--Yau three--folds have
been studied in ref.~\cite{llp}. There is, however, one important way in
which our non--zero mode differs from other non--zero
modes in pure 11--dimensional supergravity. Whereas the latter may be viewed
as an optional feature of generalized Kaluza-Klein reduction, the non--zero
mode in Ho\v{r}ava--Witten theory that we have identified cannot be turned
off. This can be seen from the fact that the constant $\a$ in
expression~\eqref{nonzero} cannot be set to zero, unlike the case in pure
11--dimensional supergravity where it would be arbitrary, since it is fixed by
eq.~\eqref{alpha} in terms of Calabi--Yau data. This fact is, of course,
intimately related to the existence of the boundary source terms, particularly
in the Bianchi identity~\eqref{Bianchi}. As we will see, keeping the non--zero
mode in the derivation of the five--dimensional action is crucial to finding
a solution of this theory that corresponds to the deformed Calabi--Yau
space discussed in the previous section.

Let us now turn to a discussion of the boundary theories. In the
five--dimensional space $M_5$ of the reduced theory, the orbifold fixed
planes constitute four--dimensional hypersurfaces which we denote by
$M_4^{(i)}$, $i=1,2$. Clearly, since we have used the standard embedding,
there will be an $E_6$ gauge field $A_\m^{(1)}$accompanied by gauginos and
gauge matter fields on the orbifold plane $M_4^{(1)}$. For simplicity,
we will set these gauge matter fields to zero in the following. The field
content of the orbifold plane $M_4^{(2)}$ consists of an $E_8$ gauge field
$A_\m^{(2)}$ and the corresponding gauginos. In addition, there is another
important boundary effect which results from the non--zero internal gauge
field and gravity curvatures. More precisely, note that
\begin{equation}
 \int_X\sqrt{\O}\, \tr F_{AB}^{(1)}F^{(1)AB} 
    = \int_X \sqrt{\O}\,\tr R_{AB}R^{AB} = -16\sqrt{2}\p v\left(\frac{4\p}{\k}
                          \right)^{2/3}\a\; ,\qquad
 F_{AB}^{(2)}= 0\; .\label{intcurv}
\end{equation}
In view of the boundary actions~\eqref{SYM}, it follows that we will retain
cosmological type terms with opposite signs on the two boundaries.
Note that the size of those terms is set by the same constant $\a$,
given by eq.~\eqref{alpha}, which determines the magnitude of the non--zero
mode. The boundary cosmological terms are another important ingredient in
reproducing the 11--dimensional background as a solution of the
five--dimensional theory.

We can now compute the five--dimensional effective action of
Ho\v rava--Witten theory. Using the field
configuration~\eqref{metric1}--\eqref{intcurv} we find from the
action~\eqref{action}--\eqref{SYM} that
\begin{equation}
 S_5 = S_{\rm grav}+S_{\rm hyper}+S_{\rm bound}\label{S5}
\end{equation}
where
\bsea
 S_{\rm grav} &=& -\frac{1}{2\k_5^2}\int_{M_5}\sqrt{-g}\left[
                  R+\frac{3}{2}\cF_{\a\b}\cF^{\a\b}+\frac{1}{\sqrt{2}}
                 \e^{\a\b\g\d\e}\cA_\a\cF_{\b\g}\cF_{\d\e}\right] \\
 S_{\rm hyper} &=& -\frac{1}{2\k_5^2}\int_{M_5}\sqrt{-g}\left[
                   \frac{1}{2}V^{-2}\partial_\a V\partial^\a V
                   +2V^{-1}\partial_\a\x\partial^\a\bar{\x}
                   +\frac{1}{24}V^2G_{\a\b\g\d}G^{\a\b\g\d}
                   \right.\nn \\
                && \left.\qquad\qquad\qquad\qquad
                    +\frac{\sqrt{2}}{24}\e^{\a\b\g\d\e}G_{\a\b\g\d}
                   \left(i(\x\partial_\e\bar{\x}-\bar{\x}\partial_\e\x )+
                   2\a\cA_\e\right)+\frac{1}{3}V^{-2}\a^2\right]\qquad\\
 S_{\rm bound} &=& -\frac{1}{2\k_5^2}\left\{-2\sqrt{2}\int_{M_4^{(1)}}\sqrt{-g}
                   \, V^{-1}\a+2\sqrt{2}\int_{M_4^{(2)}}\sqrt{-g}\,
                   V^{-1}\a\right\} \nn \\
                && -\frac{1}{16\p\a_{\rm GUT}}
                   \sum_{i=1}^2\int_{M_4^{(i)}}\sqrt{-g}\, V\tr
                   {F_{\m\n}^{(i)}}^2 \; .\label{actparts}
\esea
In this expression, we have now dropped higher-derivative terms. The
4--form field strength $G_{\a\b\g\d}$ is subject to the Bianchi identity
\begin{equation}
 (dG)_{11\m\n\r\s} = -\frac{\k_5^2}{4\sqrt{2}\pi\a_{\rm GUT}}\left\{ 
       J^{(1)} \d (x^{11})+ J^{(2)} \d (x^{11}-\pi\r )
       \right\}_{\m\n\r\s} \label{Bianchi5}
\end{equation}       
which follows directly from the 11--dimensional Bianchi
identity~\eqref{Bianchi}. The currents $J^{(i)}$ have been defined in
eq.~\eqref{J}. The five--dimensional Newton constant $\k_5$ and the
Yang--Mills coupling $\a_{\rm GUT}$ are expressed in terms of
11--dimensional quantities as~\footnote{The following relations are given
for the normalization of the 11--dimensional action as in eq.~\eqref{action}.
If instead the normalization of~\cite{conrad} is used the expression for 
$\a_{\rm GUT}$ gets rescaled to 
$a_{\rm GUT}=2^{1/3}\left(\k^2/2v\right)\left(4\p/\k\right)^{2/3}$ 
Otherwise the action and Bianchi identities are unchanged, except that in 
the expression~\eqref{intcurv} for $\alpha$ the RHS is multiplied 
by $2^{1/3}$.}
\begin{equation}
 \k_5^2=\frac{\k^2}{v}\; ,\qquad \a_{\rm GUT} = \frac{\k^2}{2v}\left(
   \frac{4\p}{\k}\right)^{2/3}\; .
\end{equation}
We have checked the consistency of the truncation which leads
to the above action by an explicit reduction of the 11--dimensional
equations of motion to five dimensions. Note that the potential terms in the
bulk and on the boundaries arise precisely from the inclusion of the
non--zero mode and the gauge and gravity field strengths, respectively.
Since we have compactified on a Calabi--Yau space, we expect
the bulk part of the above action to have eight preserved supercharges
and, therefore, to correspond to minimal $N=1$ supergravity in five
dimensions. Accordingly, let us compare the result \eqref{actparts} to the known
$N=1$ supergravity--matter theories in five
dimensions~\cite{cn,GST1,GST2,Sierra}. 

In these theories, the scalar fields in the universal hypermultiplet 
parameterize a quaternionic manifold with coset structure
$\cM_Q=SU(2,1)/SU(2)\times U(1)$. Hence, to compare our action to these
we should dualize the three--form $C_{\a\b\g}$ to a scalar field $\s$ by
setting (in the bulk)
\begin{equation}
 G_{\a\b\g\d} = \frac{1}{\sqrt{2}}V^{-2}\e_{\a\b\g\d\e}\left(\partial^\e\s
                -i(\x\partial^\e\bar{\x}-\bar{\x}\partial^\e\x )-2
                \a\cA^\e\right)\; .
\end{equation}
Then the hypermultiplet part of the action (\ref{actparts}b) can be written as
\begin{equation}
 S_{\rm hyper} = -\frac{v}{2\k^2}\int_{M_5}\sqrt{-g}\left[ h_{uv}\nabla_\a
                 q^u\nabla^\a q^v +\frac{1}{3}V^{-2}\a^2\right]
 \label{hyper}
\end{equation}
where $q^u=(V,\s ,\x ,\bar{\x})$. The covariant derivative $\nabla_\a$ is
defined as $\nabla_\a q^u= \partial_\a q^u+\a\cA_\a k^u$ with
$k^u=(0,-2,0,0)$. The sigma model metric $h_{uv} = \partial_u\partial_vK_Q$
can be computed from the K\"ahler potential
\begin{equation}
 K_Q=-\ln (S+\bar{S}-2C\bar{C})\; ,\quad S=V+\x\bar{\x}+i\s\; ,
     \quad C=\x\; .
\end{equation}
Consequently, the hypermultiplet scalars $q^u$ parameterize a K\"ahler
manifold with metric $h_{uv}$. It can be demonstrated that $k^u$ is a Killing vector
on this manifold. Using the expressions given in ref.~\cite{strom}, one can
show that this manifold is quaternionic with coset structure $\cM_Q$.
Hence, the terms in eq.~\eqref{hyper} that are independent of $\a$ describe the
known form of the universal hypermultiplet action. How do we interpret
the extra terms in the hypermultiplet action depending on
$\a$? A hint is provided by the fact that one of these $\a$-dependent terms
modifies the flat derivative in the kinetic energy to
a generalized derivative $\nabla_\a$. This is exactly the
combination that we would need if one wanted to gauge the $U(1)$ symmetry on
$\cM_Q$ corresponding to the Killing vector $k^u$, using the gauge field
$\cA_\a$ in the gravity supermultiplet. In fact,
investigation of the other terms in the action, including the
fermions, shows that the resulting five-dimensional theory is
precisely a gauged form of supergravity. Not only is a $U(1)$ isometry
of $\cM_Q$ gauged, but at the same time a $U(1)$ subgroup
of the $SU(2)$ automorphism group is also gauged. 

What about the remaining $\a$-dependent potential term in the
hypermultiplet action? From $D=4$, $N=2$  theories, we are used to the
idea that gauging a symmetry of the quaternionic manifold describing 
hypermultiplets generically introduces potential terms into the
action when supersymmetry is preserved (see for
instance~\cite{andetal}). Such potential terms can be thought of as
the generalization of pure Fayet-Iliopoulos (FI) terms. This is precisely what
happens in our theory as well, with the gauging of the  $U(1)$ subgroup
inducing the $\a$-dependent potential term in ~\eqref{hyper}.
The general gauged action will be discussed in more detail in~\cite{losw}.
Certain pure FI terms were previously considered in~\cite{GST2}, but, to our
knowledge, such a theory with general gauging has not been constructed
previously in five dimensions. 

The phenomenon that the inclusion of non-zero
modes leads to gauged supergravity theories has already been observed in type
II Calabi-Yau compactifications~\cite{sp,michelson}. From the form of the 
Killing vector, we see that it is only the scalar
field $\s$, dual to the 4--form $G_{\a\b\g\d}$, which is charged
under the $U(1)$ symmetry. Its charge is fixed by $\a$. We note that
this charge is quantized since, suitably normalized, $\tr R\w R$ is an
element of $H^{2,2}(X,{\bf Z})$. In the brane description of the
theory, this is a reflection of the fact that the five-brane charge is
quantized.

To analyze the supersymmetry properties of the solutions shortly to be
discussed, we need the supersymmetry variations of the fermions
associated with the theory~\eqref{S5}. They can be obtained either by
a reduction of the 11--dimensional gravitino variation~\eqref{susy} or
by generalizing the known five--dimensional transformations~\cite{GST1,Sierra}
by matching onto gauged four--dimensional $N=2$ theories. It is
sufficient for our purposes to keep the bosonic terms only. Both approaches
lead to
\bea
 \d \psi_\a^i &=& D_\a\e^i 
     + \frac{\sqrt{2}i}{8}
          \left({\g_\a}^{\b\g}-4\d_\a^\b\g^\g\right)\cF_{\b\g}\e^i
     - \frac{1}{2}V^{-1/2}\left(
        \partial_\a\x\,{(\t_1-i\t_2)^i}_j
        - \partial_\a{\bar\x}\,{(\t_1+i\t_2)^i}_j \right) \e^j
     \nn \\ &&
     - \frac{\sqrt{2}i}{96}V{\e_\a}^{\b\g\d\e}G_{\b\g\d\e}{(\t_3)^i}_j\e^j
     - \frac{\sqrt{2}}{12}\a V^{-1}\e (x^{11})\g_\a{(\t_3)^i}_j\e^j 
     \nn \\
 \d\z^i &=& \frac{\sqrt{2}}{48}V\e^{\a\b\g\d\e}G_{\a\b\g\d}\g_\e\e^i
     - \frac{i}{2}V^{-1/2}\g^\a\left(
        \partial_\a\x\,{(\t_1-i\t_2)^i}_j
        + \partial_\a{\bar\x}\,{(\t_1+i\t_2)^i}_j \right) \e^j
     \label{susy5} \\ &&
     + \frac{i}{2}V^{-1}\g_\b\partial^\b V\e^i
     - \frac{i}{\sqrt{2}}\a V^{-1}\e (x^{11}){(\t_3)^i}_j\e^j \nn
\eea
where $\t_i$ are the Pauli spin matrices. 

In summary, we see that the relevant five-dimensional effective theory
for the reduction of Ho\v{r}ava-Witten theory is a gauged $N=1$ supergravity
theory
with bulk and boundary potentials. While we have calculated the theory only to
order $\k^{2/3}$, one would expect that M--theory corrections can be
described in the same type of theory. For this reason, it would be very
desirable to construct the most general gauged five--dimensional $N=1$
supergravity theory coupled to general $N=1$ four--dimensional boundary
theories with vector and chiral multiplets ~\cite{losw}. In the context of global
supersymmetry, such boundary theories in five dimensions have been studied
in ref.~\cite{sharpe}. In this paper, we content ourselves with having 
identified some of the crucial generalizations that would be required.


\section{The domain--wall solution}


Let us recapitulate what we have done so far. To arrive at a simple form
for the five dimensional effective action, we have absorbed the
deformation of the Calabi--Yau background metric into the 
five--dimensional moduli. Effectively, we could then carry out the reduction
on a Calabi--Yau space but had to explicitly keep the antisymmetric tensor
part of the background as a non--zero mode in the reduction.
As a consequence, although Witten's original background preserved only
four supercharges, the effective bulk theory has twice that number
of preserved supercharges, corresponding to minimal $N=1$ supergravity in five
dimensions. For consistency, we should now be able to find the
deformations of the Calabi--Yau background as solutions of
the effective five--dimensional theory. These solutions should break half
the supersymmetry of the five--dimensional bulk theory and preserve Poincar\'e
invariance in four dimensions. Hence, we expect there to be a three--brane
domain wall in five dimensions with a worldvolume lying in the four
uncompactified directions. This domain wall can be viewed as the ``vacuum'' of
the five--dimensional theory, in the sense that it provides the appropriate
background for a reduction to the $D=4$, $N=1$ effective theory.

This expectation is made stronger if we recall the brane picture
of Witten's background. We argued that this could be
described by five--branes with equal amounts of five-brane
charge living on the orbifold planes. From the five-dimensional perspective,
the five--branes appear as three-branes living on the orbifold fixed planes.
Thus, in five dimensions, Witten's background must correspond to a pair of
parallel three-branes. 

We notice that the theory~\eqref{S5} has all of the prerequisites necessary for
such a three--brane solution to exist. Generally, in order to have a
$(D-2)$--brane in a $D$--dimensional theory, one needs to have a $(D-1)$--form
field or, equivalently, a cosmological constant. This is familiar from the
eight--brane~\cite{8brane} in the massive type IIA supergravity in ten
dimensions~\cite{romans}, and has been systematically studied for theories in
arbitrary dimension obtained by generalized (Scherk-Schwarz) dimensional
reduction~\cite{dom}. In our case, this cosmological term is provided by
the bulk potential term in the action~\eqref{S5}. From the
viewpoint of the bulk theory, we could have multi three--brane solutions with
an arbitrary number of parallel branes located at various places in the
$x^{11}$ direction. As is well known, however, elementary brane solutions have
singularities at the location of the branes, needing to be supported by
source terms. The natural candidates for those source terms, in our case, are
the boundary actions. Given the anomaly-cancelation requirements,
this restricts the possible solutions to those representing a pair of
parallel three--branes corresponding to the orbifold planes.

From the above discussion, it is clear that in order to find a three-brane
solution, we should start with the Ansatz
\bea
 ds_5^2 &=& a(y)^2dx^\m dx^\n\eta_{\m\n}+b(y)^2dy^2  \\
 V &=& V(y)\nn
\eea
where $a$ and $b$ are functions of $y=x^{11}$ and all other fields vanish.
The general solution for this Ansatz, satisfying the equations of motion
derived from action~\eqref{S5}, is given by
\bea
 a &=&a_0H^{1/2}\nn \\
 b &=& b_0H^2\qquad\qquad H=\frac{\sqrt{2}}{3}\a|y|+c_0 \label{sol}\\
 V &=&b_0H^3 \nn
\eea
where $a_0$, $b_0$ and $c_0$ are constants. We note that the boundary
source terms have fixed the form of the harmonic function $H$ in the
above solution. Without specific information about the sources, the function
$H$ would generically be glued together from 
an arbitrary number of linear pieces with
slopes $\pm\frac{\sqrt{2}}{3}$$\a$. The edges of each piece would then indicate
the location of the source terms. The necessity of matching the boundary
sources at $y=0$ and $\p\r$, however, has forced us to consider only two such
linear pieces, namely $y\in [0,\p\r ]$ and $y\in [-\p\r ,0]$. These pieces are
glued together at $y=0$ and $\p\r$
(recall here that we have identified $\p\r$ and $-\p\r$). Therefore, we
have 
\bea
\partial_y^2H &=& \frac{2\sqrt{2}}{3}\a(\d (y)-\d (y-\p\r )) 
\eea
which shows
that the solution represents two parallel three--branes located at the
orbifold planes. 

We stress that this solution solves the five--dimensional
theory~\eqref{S5} exactly, whereas the original deformed Calabi--Yau
solution was only an approximation to order $\k^{2/3}$.
It is straightforward to show that the linearized version of~\eqref{sol},
that is, the expansion to first order in $\a =O(\k^{2/3})$, coincides with
Witten's solution~\eqref{metric}, \eqref{corr} upon appropriate
matching of the integration constants. Hence, we have found an exact
generalization, good to all orders in $\k$, of the linearized 11--dimensional
solution. 

Of course, we still have to check that our solution preserves
half of the supersymmetries. When $g_{\a\b}$ and $V$ are the only non--zero
fields, the supersymmetry transformations~\eqref{susy5} simplify to
\bea
 \d\psi_\a^i &=&  D_\a\e^i -\frac{\sqrt{2}}{12}\a\,\e (y)V^{-1}\g_\a\,
                  {(\t_3)^i}_j\e^j\nn \\
 \d\z^i &=& \frac{i}{2}V^{-1}\g_\b\partial^\b V\e^i-\frac{i}{\sqrt{2}}\a\,
            \e (y) V^{-1}\, {(\t_3)^i}_j\e^j\nn\; .
\eea
The Killing spinor equations $\d\psi_\a^i =0$, $\d\z^i=0$ are satisfied
for the solution~\eqref{sol} if we require that the spinor $\e^i$ is
given by
\begin{equation}
 \e^i = H^{1/4}\e^i_0\; ,\quad \g_{11}\e^i_0 = (\t_3)^i_j\e^j_0
\end{equation}
where $\e^i_0$ is a constant symplectic Majorana spinor. This shows that
we have indeed found a BPS solution preserving four of the eight bulk
supercharges.

Let us discuss the meaning of this solution in some detail. First, we notice
that it fits into the general scheme of domain wall solutions in various
dimensions~\footnote{In the notation of ref.~\cite{dom}, it corresponds
to choosing $D=5$, $\Delta = 4/3$ and $a(5)=2$.}. It is, however, a new
solution to the gauged supergravity action~\eqref{S5} in five dimensions
which has not been constructed previously. In addition, its source terms
are naturally provided by the boundary actions resulting from
Ho\v rava--Witten theory. Most importantly, it constitutes the fundamental
vacuum  solution of a phenomenologically relevant theory. The two parallel
three--branes of the solution, separated by the bulk, are oriented in
the four uncompactified space--time dimensions, and carry the physical
low--energy gauge and matter fields. Therefore, from the
low--energy point of view where the orbifold is not resolved the 
three--brane worldvolume is identified with four--dimensional space--time.
In this sense the Universe lives on the worldvolume of a three--brane.

Although we have found an exact solution to the (lowest order) low energy
theory, thereby improving previous results, it is not clear whether the
solution will be exact in the full theory. Strominger~\cite{strom} has
argued that the
all--loop corrections (corresponding to corrections to the effective
action proportional to powers of $\k^{4/3}/V$, in our notation) to the
quaternionic metric of the universal hypermultiplet can be actually
absorbed into a shift of $V$, so that the metric is unchanged. This
implies that our solution would be unaffected by such corrections. On the
other hand, we have no general argument why the solution should be
protected against corrections from higher derivative terms. 

In any case, we believe, that pursuing the construction of
five--dimensional gauged supergravities with boundaries, and the analysis
of their soliton structure, in the way indicated in this paper
might provide important insights into low energy particle phenomenology
as well as early universe cosmology.

\vspace{0.4cm}

{\bf Acknowledgments} 
A.~L.~is supported in part by a fellowship from Deutsche
Forschungsgemeinschaft DFG). A.~L.~and B.~A.~O.~are supported in part by 
DOE under contract No. DE-AC02-76-ER-03071. D.~W.~is supported in part by
DOE under contract No. DE-FG02-91ER40671. 



\end{document}